\documentclass[fleqn,10pt]{wlscirep}
\usepackage[utf8]{inputenc}
\usepackage[T1]{fontenc}
\usepackage{comment}
\usepackage{amsmath, tabularx}
\usepackage{bigstrut}
\usepackage{nccmath}
\usepackage[sort, numbers]{natbib}
\usepackage{dirtytalk}
\usepackage{multirow}
\title{From Labels to Priors in Capsule Endoscopy: A Prior Guided Approach for Improving Generalization with Few Labels}
\author[1,*]{Anuja Vats}
\author[1,2]{Ahmed Mohammed}
\author[1]{Marius Pedersen}

\affil[1]{Department of Computer Science, NTNU, Gjøvik, 2819, Norway}
\affil[2]{SINTEF Digital, Smart Sensor Systems, Oslo, Norway}

\affil[*]{anuja.vats@ntnu.no}


\begin{abstract}
The lack of generalizability of deep learning approaches for the automated diagnosis of pathologies in Wireless Capsule Endoscopy (WCE) has prevented any significant advantages from trickling down to real clinical practices. As a result, disease management using WCE continues to depend on exhaustive manual investigations by medical experts. This explains its limited use despite several advantages. Prior works have considered using higher quality and quantity of labels as a way of tackling the lack of generalization, however this is hardly scalable considering pathology diversity not to mention that labeling large datasets encumbers the medical staff additionally. We propose using freely available domain knowledge as priors to learn more robust and generalizable representations. We experimentally show that domain priors can benefit representations by acting in proxy of labels, thereby significantly reducing the labeling requirement while still enabling fully unsupervised yet pathology-aware learning. We use the contrastive objective along with prior-guided views during pretraining, where the view choices inspire sensitivity to pathological information. Extensive experiments on three datasets show that our method performs better than (or closes gap with) the state-of-the-art in the domain, establishing a new benchmark in pathology classification and cross-dataset generalization, as well as scaling to unseen pathology categories.  

\end{abstract}
\begin{document}
\flushbottom
\maketitle
%
%
\thispagestyle{empty}

\section*{Introduction}

\label{sec:intro}
WCE has become indispensable for the diagnostic inspection and management of gastrointestinal diseases. The rise in its preference among clinicians and patients is due not only to being minimally intrusive, but also for allowing more comfortable outpatient procedures, reducing the need for unnecessary hospital admissions. However, the increase in its uptake as an alternative to traditional endoscopy comes with an overhead, large volumes of post-procedure data to be examined by clinicians. Computer-Aided Diagnosis (CADx) has shown promise in various clinical applications including WCE by automating cumbersome aspects of the traditional diagnostic pipeline as well as supplementing diagnoses with second objective opinions \cite{YANASE2019CADXreview}. However, in WCE, despite progress in imaging, reliable CADx remains largely unaddressed \cite{hwang2018application}. Most deep learning based approaches for automatic detection and classification in WCE are fully supervised. The generalization ability for such supervised objectives critically depends on the size and diversity of samples in each class. Since procuring labels adequately enough to emulate real clinical scenarios with multiple pathologies is highly time and resource intensive, most studies in WCE focus on a particular class of abnormality such as polyps \cite{mohammed2018polyp,saito2020protrudinglesion}, angiectasia \cite{leenhardt2019angiectasia}, hookworm \cite{tmi_hookworm}, bleeding \cite{aoki2020blood} etc. Multi-pathology classification in WCE continues to be challenging even with supervised approaches and millions of images \cite{ding2019gastroenterologist,vats2021learning}. This can be partially attributed to the challenges particular to this domain. A WCE image can be a mix of different elements like bubbles, fluids, local anatomy (shape, color, textures), varying illumination, etc. (these elements are hereby referred to as factors in the paper). Factors that characterize abnormalities like color, texture, scale, etc. exist locally with other normal factors like varying local anatomy, texture associated with normal tissue, constituents like gastrointestinal fluid, bubbles, food remnants, etc. and since all these factors interplay within a single frame, extracting reliable pathology-features for their classification has been a long-standing challenge in WCE \cite{ding2019gastroenterologist}. 
The need for generalization of CADx to multiple pathologies in WCE, but with accessability only to unlabelled data, motivates this approach for learning under complete unsupervision.


In the unsupervised paradigm, the core approach is that of discovering a sufficiently generalizable representation space corresponding to the unlabeled images such that the resulting space encodes information while exhibiting certain beneficial properties (properties discussed in detail in Sec. \ref{sec:embeddingsec}). 
Multi-view contrastive learning has recently become a powerful component for unsupervised representation learning \cite{oord2018cpc,wu2018unsupervised,he2020moco,misra2020self,NEURIPS2020_tian,tian2020cmc,chen2020simclr}. Here, in the absence of labels, representations are learned by maximizing the information shared between two views/crops of the same image \cite{tian2020cmc}. This translates to developing invariances (for factors not shared between the views) alongside feature extraction (for shared factors) and is guided by the choice: \say{what is mutual} between the views.
Recent work by Tian et al. \cite{NEURIPS2020_tian} argues in favor of maximizing mutual information (MI) but selectively such that only the information of consequence is shared but no more. What this means for most large-scale datasets like ImageNet \cite{krizhevsky2017imagenet}, STL-10 \cite{coates2011stl10} etc. employed in contrastive pretraining is that since most images consist of one primary instance (for example one or many cars), the features of interest can be safely assumed to co-occur in two random crops or remain as the dominant factor in two randomly augmented versions of the image. By minimizing the distance between these crops in the latent space, features for mutual factors (car and its parts) are extracted and invariances to non-mutual factors (like background trees) are developed.

This assumption of mutual co-occurrence of instances may not hold true for some domains including WCE where multiple small-scale factors are prevalent. Random augmentations to create two views in such domains may share too much (uninteresting factors prevail in one or both crops) or too little (small scale pathology missing from one or both crops). Moreover, medical datasets as used in this work, may come from a true-unlabeled data corpus, with no a priori information on the samples per class or even the number of classes inherent in the data. In the absence of annotations and more so adequate representation of abnormality types and subtypes, an approach to tune the representations such that they can be preferential in attending to certain factors e.g. pathologies while exhibiting invariance to others e.g. normal variations, is the primary objective of this work. In keeping with the strict assumption of \say{no labels for pretraining} we propose an unsupervised approach that exploits simple domain priors for learning progressively selective representations. The main contributions of this work are:




\begin{itemize}
    \item We propose an approach to reduce the labeling requirement through the use of simplistic domain priors to guide representation learning.
    \item We propose a new type of negative for contrastive learning: Within-Instance Negative (WIN).
    \item We present a framework (first to the best of our knowledge) for WCE classification that generalizes across datasets and to new pathology classes (including diseases with low prevalence). Tables \ref{tab:LE} and \ref{tab:fft} present multi-dataset pathology classification benchmarks to further improve the ease of comparison of cross-dataset generalization in the field. In addition, the pretrained weights (to be publicly released) can be used for the creation of weak-labels on a number of WCE pathologies and capsule modalities. (weak labels can be seen in Figure \ref{ScoreCam}(b.) and supplementary video.)
    \item Our method surpasses traditional transfer-learning approaches such as pre-training on much larger ImageNet dataset in accuracy by 1.6\% (94.7-93.1), as well as other recent works in multi-class classification by 1.4\% \cite{valerio2019GIANA} and 1.53\% \cite{guo20203class}  on CAD-CAP dataset, (refer Table \ref{tab:fft}). We also surpass random-augmentation based contrastive learning \cite{misra2020self}, indicating that prior-guided views are superior to random augmentation based views.
\end{itemize}


\section{Related work}
\textbf{Wireless Capsule Endoscopy}: Since polyps are precursors for cancer, the majority of CADx approaches in both colonoscopy and WCE have focussed on its detection in images \cite{fan2020pranet_NSR, li2009oldest,mohammed2018polyp}. With the popularity of convolutional neural network based feature extractors, other pathologies like bleeding, erosions, ulcerations, angiectasia \cite{leenhardt2019angiectasia,aoki2020blood,BIN_WCE} etc. have been increasingly included for automated diagnosis. However, all these approaches either tackle only one of many pathologies or consider all pathologies as a single class. A few approaches attempt multi-pathology classification \cite{vats2021learning, valerio2019GIANA} using fully supervised approaches. These approaches typically use pretrained networks to compensate for overfitting on smaller WCE datasets. Surprisingly, even with large scale WCE datasets \cite{ding2019gastroenterologist} multi-pathology classification remains challenging. Earlier works \cite{guo20203class,vats2021learning} have identified the diverse characteristics in WCE images to be the challenge. Recently, a semi-supervised approach \cite{guo20203class} performs multi-pathology classification using a combination of labeled and unlabeled data with dilated convolutional layers for attending  to abnormal regions. In this work, we propose to expand this quest and pretrain with unlabeled data only, for multi-pathology classification.\\
\textbf{Self supervised learning}: The supervision in self-supervision comes from exploiting signals inherent within the data. The signal may come from predicting missing, corrupted, or future information or by establishing correspondence between inputs  \cite{noroozi2016jigsaw,gidaris2018rotnet,holmberg2020self_NATURE} or even from comparison between multiple views of an input image \cite{wu2018unsupervised,tian2020cmc,chen2020simclr,oord2018cpc}. In this multi-view contrastive learning, supervision arises from mapping two different views of an image, close together in the latent space. Although carefully constructed views have proven to be the key to good performance in downstream tasks like classification \cite{chen2020mocov2} in the natural domain, questions about its sufficiency have already been raised as newer domains are considered. In \cite{NEURIPS2020_tian} the authors discuss challenges for multi-factor datasets in the context of natural images and propose using a subset of labels during pretraining to selectively tune for desirable factors. Another recent work \cite{ContrastiveCrop} proposes initializing views randomly but using approximate heatmaps from higher level convolutional layers as guidance for curating better views adding a little computation overhead. In this work, we investigate how simple prior-knowledge from the domain can be used to do the same without additional components to the contrastive pipeline, enabling fully unsupervised but still guided pretraining.

\begin{figure}[t]
\center
\includegraphics[width=6cm,height=5cm]{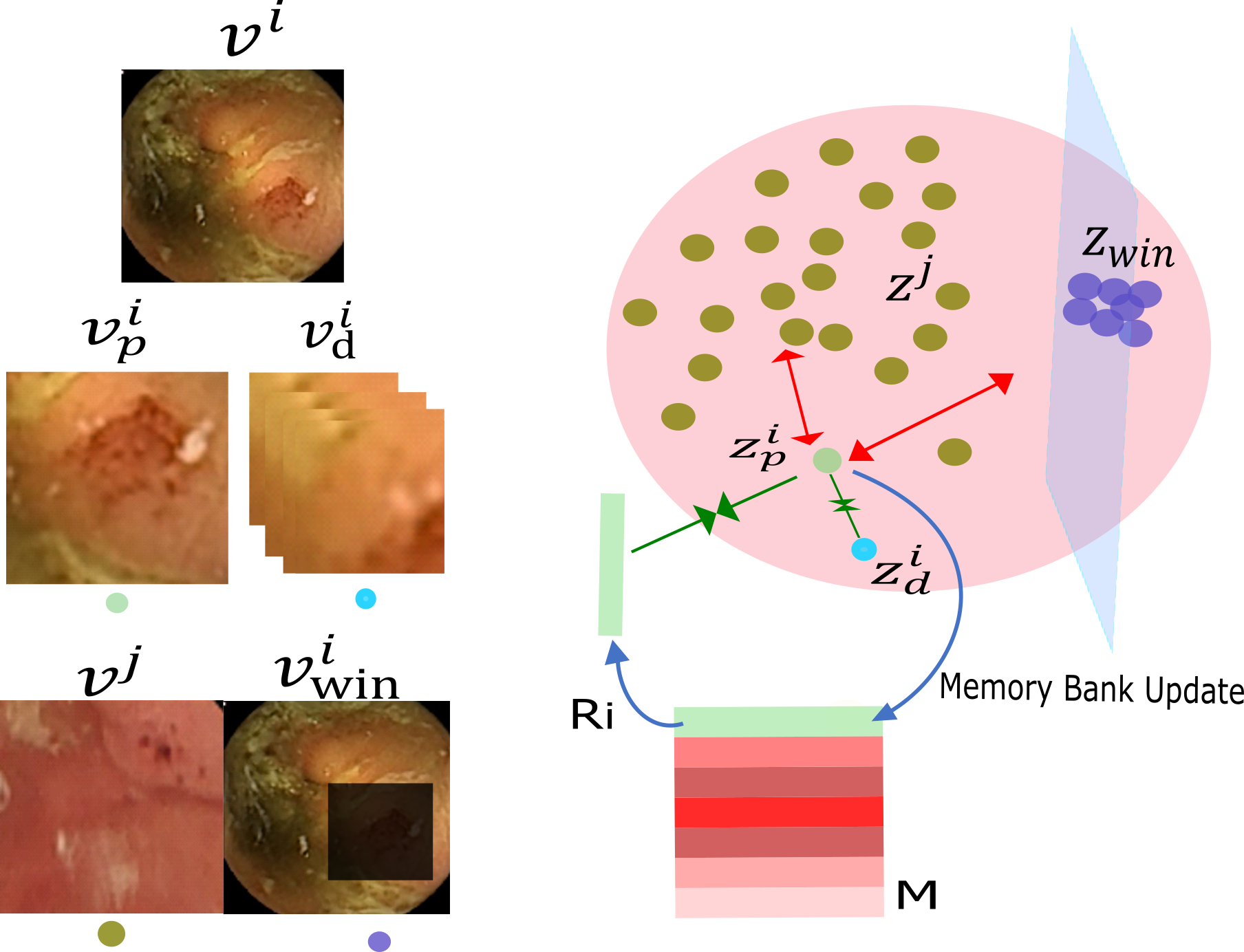}
\caption{Proposed approach: Given an unlabeled WCE image \(v^i\), we use priors to create special views namely a pathology-aware view \(v_p^i\), a pathology-ignorant view \(v_{win}^i\) and a distorted view \(v_d^i\). \(z_*^*\) denotes the encodings of these views. We use combinations of these views with contrastive objectives to strategically emphasize on pathology features during training. The resulting feature space (pink circle) shows how pathology-aware features \(z_p^i\) and \(z^j\) dominate the output space and push away from pathology-ignorant features \(z_{win}\). We show the real contrastive space and analyze its properties fig.\ref{evolution} and sec.\ref{sec:embeddingsec}.
}
\label{Views}
\end{figure}

\section{Methodology}

As stated above, we propose the use of simplistic domain priors to guide the learning to be selective to pathology factors in an image during pretraining. The priors should be such that the scale and attention of features attunes automatically to even slight, obscure signs of pathology within each image much like a selectivity filter, that activates more for some parts from multiple foreground objects. Since priors are key to achieving this, in favor of high generalizability (to many pathology classes even with low prevalence and irrespective of capsule modalities), the priors must be overly general. The priors are as follows:\\
\textbf{ Redness Prior : } This prior is based on a generic property that many pathologies are known to exhibit. It is shown that the presence of many pathologies tends to locally influence the appearance by exhibiting an increased level of redness along with other traits that may be pathology-specific \cite{mizukami2017redness,mcnamara2019erythematousred}. The pathology-specific traits may be vastly varying, and many pathologies may be identified by means of other more conclusive traits. However, an area with increased redness can be associated with a high probability of occurrence of the abnormality itself or associated traits. We use redness to serve as an initial proxy to shift the scale and attention toward pathology-specific traits within an image (Section \ref{sec:PGCON}).\\
\textbf{ Locality Prior : } Datasets with multiple local factors, such as in WCE, have an achilles heel that we exploit to create a new Within-Instance Negative (WIN). As a pathology can often be obscure and localized, it is surrounded by normal tissue and its variations (normal mucosal folds, color, texture, vascular pattern, bubbles, etc.). Some pathologies like a polyp may share the same visual properties as the region it occurs in (texture, color, etc.), but despite the similarity, the actual polyp is local. We use the locality prior to inspire that different local regions within an image must weight differently. We do this through a new type of negative (WIN) for within-instance contrast (Section \ref{sec:WIN}).

\begin{figure*}[!thbp]
\centering
\includegraphics[width=15cm, height=4.5cm]{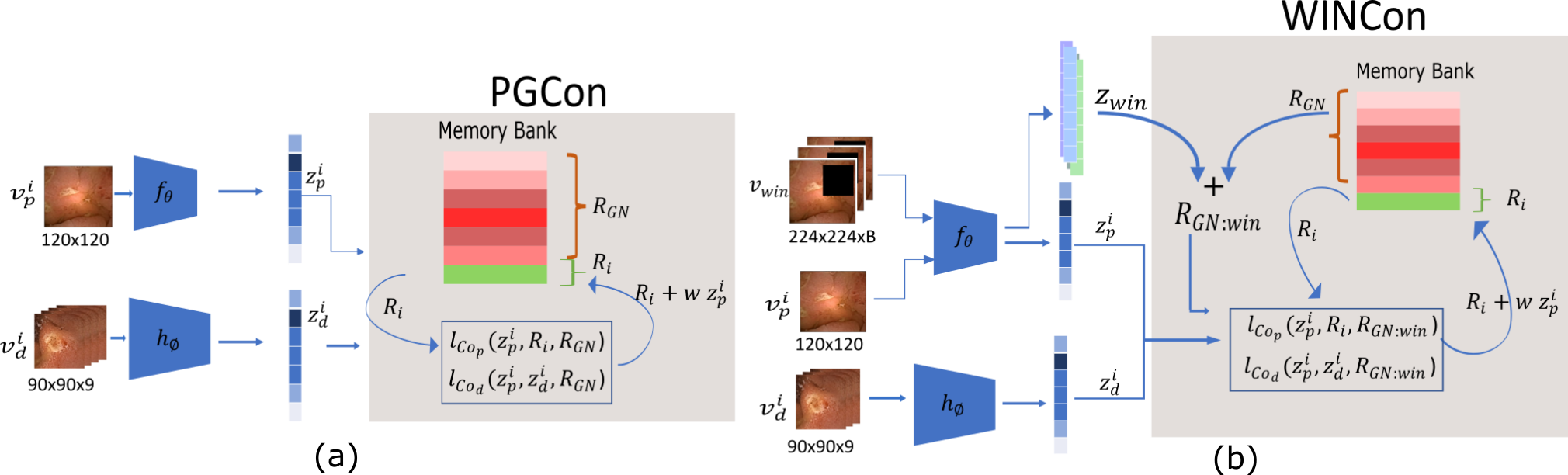}
\caption{Overview of proposed objectives. (a) Two views $v_i^p$ (prior view) and $v_i^d$ (distorted view) constructed from the same image are encoded as $z_p^i$, $z_d^i$ respectively. The contrastive objective uses these as well as representations $R_{GN}$ and $R_i$  from an evolving memory bank \(\mathcal{M}\) to minimize distance between positives and maximize distance between negatives. (b) In addition to $v_i^p$ and $v_i^d$, WINCon uses $v_{win}$ derived from all images in the batch (B) by removing regions suspected of pathology. These $v_{win}$ are transformed into $z_{win}$ and used as additional negatives. Refer to sections \ref{sec:PGCON} and \ref{sec:WIN} for more details.}
\label{fig_sim}
\end{figure*}

\subsection{Prior Guided Contrast (PGCon)} 
\label{sec:PGCON}

We use the redness prior to guide a shift in the scale (global to local) and attention (normal to pathology) of features such that they get progressively more selective to the pathology and its variation. We do this by creating two views for an image \(i\) :\\
(a.) prior-view (\(v_p^i\)) -  based on the redness prior, we extract this view by cropping a fixed square centred around the highest value pixel in the $a^*$-channel (red-green) from the CIELAB space. \(v_p^i\) is our main view and is only a small patch of the entire image, suspected of containing pathology traits. It undergoes random transformation sampled from a set of transforms \(\mathcal{T}_p\) (details in Supplementary 1.)\\
(b.) distorted-view (\(v_d^i\)) - main purpose of this view is to encourage invariance, while keeping pathology information mutual with \(v_p^i\). The factors to be invariant to are an essential consideration in choosing those views that discard irrelevant details, and this is what \(v_d^i\) helps achieve. Like natural images, even medical representations \cite{sowrirajan2020mococxr} benefit from invariance to low-level image transformations (random crop, color jitter, etc.), but such invariance in isolation fails to account for domain-related nuisance factors (for e.g., varying normal morphology, dynamic capsule orientation, floating residues, bubbles and varying anatomy of each segment of the intestine \cite{laiz2019using}). Such irrelevant variations being more pervasive than small scale abnormalities seep through in the feature space, as reported in previous literature \cite{ding2019gastroenterologist,vats2021learning}. A \say{good} representation must exhibit invariance to low-level transformations \textit{as well as} such irrelevant domain variations, thus \(v_d^i\) is a jigsaw puzzle composed of nine tiles from the input image. The transformation for tiles common with \(v_p^i\) are sampled from  \(\mathcal{T}_p\), the same as \(v_p^i\). In this way, randomly augmenting the shared tiles between the two views using transforms sampled from the same set \(\mathcal{T}_p\), promotes transformation invariance. The other tiles (those not shared between views), suspected of exhibiting irrelevant factors, are purposefully distorted using transforms sampled from another set \(\mathcal{T}_d\) (details in Supplementary 1.). As there is no incentive for learning these non-mutual tiles, this promotes invariance to specific domain variations. Fig. \ref{Views} illustrates the two views.\\
\textbf{Objective Function:}
\label{Obj_PGCon}
Let \(D\) be a dataset consisting of \(n\) instances, \(\mathcal{D} = \{x_1, x_2, ..., x_n\}\), such that given an instance  \(x_i \sim \mathcal{D}\), two views \(v_p^i\) and \(v_d^i\) can be constructed. Convolutional encoders \(f\) and \(h\), parametrized by \(\theta\) and \(\phi\) respectively,  non-linearly transform these input views such that  \(z_p^i=f_\theta(v_p^i) \) and \(z_d^i = h_\phi(v_d^i)\) in a 128-dimensional vector space.  Additionally, let a memory bank \(\mathcal{M}\) of size \(n\) accumulate an exponential moving average for each \(f_\theta(v_p^i) \) from all previous iterations of seeing image \(x_i\), this moving average is given by $R_i$. Since only a few $R_i$s (limited to batch-size) are updated at each iteration from \(\mathcal{M}\), \( R_i\) is a stale version of \(z_p^i\) that will be updated in the current iteration. Fig. \ref{fig_sim}(a) illustrates the PGCon objective.

We use InfoNCE loss \cite{oord2018cpc,tian2020cmc,wu2018unsupervised,he2020moco,chen2020simclr} understood as maximizing the lower bound on mutual information \(\mathcal{I(\cdot)}\), between two pairs: \( (z_p^i, R_i)\) and \((z_p^i, z_d^i)\).
Knowing that each of \(z_p^i, z_d^i\)  and \(R_i\)  are encoded from views of \(x_i\), we denote a sample from each joint distribution \(p(z_p^i, R_i)\)  and  \(p(z_p^i, z_d^i)\)  as \textit{positive}, and a set of \(2k\) independently drawn samples from the product of marginals  \(p(z_p^i)p(R_j)\) as \textit{negatives},where \(R_j\) \(\in\) \(R_{Global Negatives}\) (\(R_{GN}\) in fig. \ref{fig_sim}) and \(j \neq i\).  The contrastive learning problem tries to optimize an objective \(\mathcal{L}_{Co}\), such that, given the pairs sampled from the two distributions, a scoring function \(g\) (equation \eqref{eq:9})  discriminates between samples from the two distributions (assigning a higher probability to samples from the joint than from the product of marginals). This can be formulated as the cross-entropy loss between the similarity scores of the positive and negative pairs as

\begin{tabularx}{\linewidth}{@{}XX@{}}
\begin{equation}
\label{eq:1}
\mathcal{L}_{Co_p} = -\mathbb{E} \: log   \left[ \frac{exp \frac{s(z_p^i,R_i)}{\tau}}{exp\frac{s(z_p^i, R_i)}{\tau} + \sum_{j \neq i} exp \frac{s(z_p^i, R_j)}{\tau}}\right]
\end{equation}
&
\begin{equation}
\label{eq:2}
\mathcal{L}_{Co_d} = -\mathbb{E} \: log   \left[ \frac{exp \frac{s( z_p^i,z_d^i)}{\tau}}{exp \frac{s(z_p^i,z_d^i)}{\tau} + \sum_{j \neq i} exp \frac{s(z_p^i,R_j )}{\tau}}\right]
\end{equation}
\end{tabularx}

The expectation in \(\mathcal{L}_{Co_p} \) is over all positive pairs  \(\{z_p^i, R_i\}_{i=1}^n\)  from the joint distribution \(p(z, R)\) and likewise for \(\mathcal{L}_{Co_d}\). \(\tau > 0\) is a scalar temperature hyperparameter. The k-negatives: \({\{R_j\}}_{j \neq i}\) are randomly retrieved from \(\mathcal{M}\) and are the encodings of prior-views \(v_p\) of other instances. These negatives are advantageous for two reasons. (a.) Over time, sampling \({R_j}\) from \(\mathcal{M}\) automatically allows mining harder negatives as over the course of training, the memory bank representations \(\{{R_j}\}_{j=1}^n\) specialize in local pathological regions and discriminating between \((z_p^i, R^j )\) amounts to discriminating between finer features that characterize pathologies. (b.) When retrieving negatives from \(\mathcal{M}\) instead of the input batch, the number of negatives gets decoupled from the batch size. 
The total objective is a weighted sum of losses in Equations \eqref{eq:1} and \eqref{eq:2} with scalar parameters \(\alpha\ =\beta = 0.5\), shown in equation \eqref{eq:3}. The scoring function \( g\) (equation \eqref{eq:9})models the ratio of densities between the joint and the product of marginals where \( s\) is the commonly adopted cosine similarity score. 
\begin{ceqn}
\begin{align}
  \label{eq:3}
 \mathcal{L}_{co} = \alpha \mathcal{L}_{Co_p} + \beta \mathcal{L}_{Co_d}
\end{align}
\end{ceqn}

\begin{equation}
\label{eq:9}
\qquad \qquad \qquad \qquad \qquad \qquad \qquad \qquad g = exp(s(z_i, R_j)) =  exp(\frac{z_i^T.R_j}{\Vert{z_i}\Vert \cdot \Vert{R_j}\Vert})
\end{equation}

\subsection{Within Instance Negative (WINCon)} 
\label{sec:WIN}
PGCon shifts attention within images so that the resulting representations encode local regions corresponding to \(v_p^i\) that is in simultaneous contrast with other \say{suspected} pathology instances in the dataset retrieved from \(\mathcal{M}\), i.e., it's global negatives. Since most common abnormalities have local prevalence, we suspect the representations to also benefit from contrasting with the leftover image, after extracting \(v_p^i\). \(v_i^{win}\) (refer Fig. \ref{Views}) is created by zeroing the pixels of \(x_i\) that form \(v_p^i\) and applying \(\mathcal{T}_{win}\) (Supplementary 1.) to the resulting image. We observe \( v^{win}\)s to be hard negatives at initial epochs (explained later in Fig. \ref{evolution}). This may be due to the continuity of normal or pathological patterns at the boundaries, as well as other possible visual similarities. Nevertheless, contrast with such WINs enable us to induce the idea that one local region within an image could be different from the rest despite boundary similarities. We verify this in the next objective: WINCon. \\

\textbf{Objective Function:}
Mathematically, it amounts to appending \(\{z_{win}\}_{i=1}^{B}\), where  \(z_{win}=f_\theta(v_{win})\) and B is the batch size to the list of global negatives in Equations \eqref{eq:1} and \eqref{eq:2}. 
The new formulation is given as :

\begin{tabularx}{\linewidth}{@{}XX@{}}
\begin{equation}
\label{eq:5}
\mathcal{L}_{Co_{p}} = -\mathbb{E} \: log   \left[ \frac{exp \frac{s(z_p^i,R_i)}{\tau}}{exp\frac{s(z_p^i, R_i)}{\tau} + \sum_{j \neq i} exp \frac{s(z_p^i, [R_j:z_{win}])}{\tau}}\right]
\end{equation}
&
\begin{equation}
\label{eq:6}
\mathcal{L}_{Co_{d}} = -\mathbb{E} \: log   \left[\frac{exp \frac{s( z_p^i,z_d^i)}{\tau}}{exp \frac{s(z_p^i,z_d^i)}{\tau} + \sum_{j \neq i} exp \frac{s(z_p^i, [R_j:z_{win}])}{\tau}}\right]
\end{equation}
\end{tabularx}

The total objective is a weighted sum of losses in \eqref{eq:5} and \eqref{eq:6}, with \(\alpha\ =\beta = 0.5\). Fig. \ref{fig_sim}(b) illustrates WINCon objective.
\begin{ceqn}
\label{eq:7}
\begin{align}
\mathcal{L}_{co_{win}} = \alpha \mathcal{L}_{Co_{p}} + \beta \mathcal{L}_{Co_{d}}
\end{align}
\end{ceqn}

\section{Datasets}
\subsection{Datasets for Training}
For contrastive pretraining, we use two different datasets: PS-DeVCEM dataset and OSF-Kvasir-Capsule dataset.\\
\textbf{PS-DeVCEM Dataset:} PS-DeVCEM data is a subset of a private capsule dataset from Pillcam Colon2, Medtronic\textsuperscript{\texttrademark} with video level labels and has previously been used in another study \cite{mohammed2020MARKDATA}. There are no image labels in the dataset. We gained this data with consent of the original authors and have used the label information only to check that the cleanliness level of the bowel is satisfactory so that videos with extremely low visibility of the muscosa are avoided as well as there is at least some abnormality in the selected study. No other information about the type/severity or prevalence of the pathology in the video has been considered. The PS-DeVCEM data consists of 80,946 images from short video segments of 12 examinations with normal and abnormal frames. Of these, a significant number of intermittent frames may comprise normalcy typically observed between episodes of gastrointestinal abnormality. The exact number of frames with pathologies as well as the classes is not known, and this corresponds to a true unlabeled setting. PS-DeVCEM data forms approximately 96\% of total pre-training data. 

 \begin{table*}[!htbp]
\renewcommand*{\arraystretch}{1}
\resizebox{1\textwidth}{!}{

  \centering
    \begin{tabular}{lllll}
    \hline
    Dataset & Classes & Image size & \multicolumn{1}{l}{\# Total images used} & \multicolumn{1}{l}{\(v_p^i\) size (train datasets only)} \bigstrut \\
    \hline
    PS-DeVCEM  & Unknown & 576x576 & 80,946 & 150 \bigstrut\\
    OSF-Kvasir & Angiectasia, Blood, Erosion, Erythematous, Lymphoid Hyperplasia, Polyp , Ulcer & 336x336 & 3498 & 100 \\
    KID  & Vascular, polypoid, Lymphangiectasia, inflammation, Aphtha, Bleeding, Chylous & 360x360 & 77 &  --\\
    KID2  & Inflammatory, Normal, Polypoid, Vascular & 360x360 & 142 & --\\
    CAD-CAP  & Inflammatory Lesion, Vascular Lesion, Normal & 576x576 & 1812 & -- \bigstrut\\
    \hline
    \end{tabular}%
    }

  \caption{Dataset Details. Attribute \(v_p^i\) size indicates the initial crop size of \(v_p^i\) for each dataset. \(v_p^i\) is resized to 120x120 irrespective of the dataset before input to the network.}
  \label{tab:dataset}%
\end{table*}%

\textbf{
OSF-Kvasir Dataset :} OSF-Kvasir-Capsule dataset \cite{smedsrud2021kvasir} with 3478 images from seven classes taken with the capsule modality Olympus EC-S10\textsuperscript{\texttrademark}. The original dataset is composed of 14 classes, out of which six (Pylorus, Ampulla of Vater, Ileocecal Valve, Normal, Reduced view Mucosa and foreign body) do not correspond to pathological findings and hence have been removed. In the remaining 8 classes, OSF-Kvasir dataset is imbalanced with just 12, 55 and 159 samples in the classes hematin, polyp, and lymphangiectasia respectively, as opposed to 866, 854 samples for other classes. In an attempt to balance the dataset, only class \say{hematin} with 12 samples has been removed, and the remaining seven classes have been used for both pretraining and downstream testing. 

\subsection{Datasets for Evaluation}
In addition, three other datasets have been used solely for evaluation, these are:\\
\textbf{Few shot KID (FS-KID) :} KID dataset \cite{koulaouzidis2017kid,pmid25088924kid1}, capsule modality MiroCam, IntroMedic\textsuperscript{\texttrademark}. FS-KID is used for few-shot classification as it comprises of a total of 77 images from seven categories with as few as five samples in some classes.\\
\textbf{Few shot KID2 (FS-KID2) :} KID2 dataset \cite{koulaouzidis2017kid,pmid28580415kid2} capsule modality MiroCam, IntroMedic\textsuperscript{\texttrademark}, similar to the OSF-Kvasir Dataset, non-pathological classes (ampulla of vater and normals from the esophagus and stomach) have been removed from FS-KID2. The final dataset consists of four classes (inflammation, normal, polypoid, and vascular) that has been class balanced (between 17 and 22 samples in each class). This is also used for few-shot classification.\\
\textbf{CAD-CAP Dataset :} CAD-CAP is a balanced dataset with 1812 images in three classes (inflammatory lesion, vascular lesion, and normal) as part of the GIANA Endoscopic Vision Challenge 2018. We use this dataset for evaluation to facilitate comparison with other works on pathology classification \cite{valerio2019GIANA, vats2021learning}.
The split for all datasets except CAD-CAP follows a 60:40 train-val split (due to very few samples in a few classes, some with just two). In CAD-CAP an 80:10:10 train-val-test split is used as the dataset is balanced and sample-sufficient in each class and traditional 1\%, 10\% and 100\% label-subsets are evaluated. The CAD-CAP val-set is used for finding the optimum epoch for test/inference on all datasets. We observed a consistent epoch-accuracy behavior for each subset of data (100 epochs-1\%, 200 epochs-10\%, 300 epochs - 100\% even between different datasets). Once this was fixed, the reported accuracy is averaged over three runs for the checkpoint with the best validation accuracy for OSF-Kvasir, FS-KID and FS-KID2, for CAD-CAP dataset we report the test accuracy. Table \ref{tab:dataset} summarizes the additional dataset details.

\section{Experiments} 
We evaluate PGCon and WINCon by transferring them to pathology classification tasks under different policies on four datasets. Through this we test cross-dataset and cross-capsule modality transfer including generalization to new, unseen pathology categories like apthae, chylous, inflammatory lesion, etc. We compare against ImageNet pretrained R50, ImageNet pretrained Densenet161 (a significantly bigger architecture), random augmentation based pretraining in PIRL\cite{misra2020self}, and recent fully and semi-supervised approaches for pathology classification \cite{valerio2019GIANA,guo20203class}.\\
\textbf{Training details: }We perform contrastive pretraining on ResNet-50 encoder (R50) \cite{he2016RESNET} with same architecture as PIRL\cite{misra2020self} for ease of comparison, with proposed loss and views (using \(\mathcal{T}_d\),\(\mathcal{T}_p\), \(\mathcal{T}_{win}\)). To obtain \(z_p^i\) we pass \(v_p^i\) through the R50 encoder up to the global average pooling layer followed by a 128-dim fully-connected (fc) layer. For \(v_d^i\) after obtaining 128-dim embeddings for each of the nine tiles, we use an additional fc-layer to produce a compressed 128-dim \(z_d^i\). We train with batch-size 64, negatives 2k=400 (k per loss term) and 600 epochs across all pretraining experiments. Similarly to \cite{misra2020self}, we use the mini-batch SGD optimizer. The learning rate schedule is cosine annealing with an initial and final value of 0.012 and 1.2x \(10^{-5}\), we start from 0.012 for all experiments, including our adaptation of PIRL to WCE. Temperature \(\tau\) is fixed at 0.07 for all experiments. 

\subsection{Task 1: Zero Shot Image Classification }
\textit{Setup:}
It is argued that transferability in large-scale feature extractors comes from the knowledge of concepts that lend easy adaptation in the face of new categories and tasks such that few examples are sufficient for generalization. We investigate, if such fundamental knowledge exists, which in our domain translates to visual concepts relating to pathologies, then there may inherently exist weak discrimination within the feature space. To test cross-dataset and new category generalization, we use the kNN based clustering approach also used in \cite{wu2018unsupervised} on CAD-CAP data (not used in pretraining). Since the evaluation occurs directly with contrastive pretraining weights derived from other datasets, with no additional labels from CAD-CAP, we call this zero-shot classification. The same setup as \cite{wu2018unsupervised} is adapted with \(\tau = 0.1\) and number of top-neighbors \(k=290\). 
\textit{Observations:} As in Table \ref{tab:knn}, our fully unsupervised performance surpasses the supervised performance in \cite{vats2021learning} by 9\% and unsupervised PIRL performance \cite{misra2020self} by almost 13\% indicating that the representations are robust enough to directly discriminate between classes of pathology without requiring any fine-tuning on target dataset .

\begin{table}[!htbp]
 
  \centering
   \resizebox{0.45\textwidth}{!}{
    \begin{tabular}{|p{19.165em}|l|}
    \toprule
    \multicolumn{2}{|p{26.83em}|}{Unsupervised Clustering on Out of Distribution Dataset} \\
    \midrule
    PIRL  & 55.0 \\
    \midrule
   Vats et al. (supervised) \cite{vats2021learning} & 58.7 \\
    \midrule
    PGCon   & \textbf{68.0} \\
    \midrule
    WINCon & 66.0 \\
    \bottomrule
    \end{tabular}
    }
  \caption{Top-1 accuracies using weighted kNN-classifier over out-of-distribution samples from CAD-CAP data to obtain an estimate for inherent discrimination and generalization to new classes. Clustering is performed over normalized 128-dimensional \(z_p^i\) embedding from R50 encoder. To our surprise, PGCon and WINCon outperforms fully supervised classification on CAD-CAP data \cite{vats2021learning}.}
   \label{tab:knn}
\end{table}%

\subsection{Task 2: Downstream Linear Classification  }
\textit{Setup:} Next we perform linear evaluation where the encoder weights are frozen and only the linear classification layers are trained. Table \ref{tab:LE} presents the first of such evaluation in WCE across all four datasets for pathology classification. As discussed earlier, the evaluation on FS-KID and FS-KID2 is few-shot due to few samples per class and in CAD-CAP we explicitly test few-shot with 1\% and 10\% label subsets. \\
\textit{Observations:} We consistently perform better than PIRL \cite{misra2020self} across all datasets and label-subset regimes with a difference in accuracy of up to 18\% observed for CAD-CAP-1\% (2 to 3 samples per class). In comparison with ImageNet pretraining, we close the gap in OSF-Kvasir, FS-KID, FS-KID2 and CAD-CAP datasets. As earlier, the improvement over Imagenet pretraining (1 million natural images against 80k unbalanced, roughly 12 times less) is more discernible in low data regimes (FS-KID, FS-KID2, CAD-CAP-1\%). We believe this to be due to our representations encoding concepts of pathologies, even before fine-tuning, that are superior to out-of-domain representations like those from Imagenet. 


\begin{table}[htbp]
 \renewcommand*{\arraystretch}{1}
  
  \centering
    \begin{tabular}{|c|cccccc|}
    \hline
\cline{2-7}    \multicolumn{1}{|c|}{} & \multicolumn{1}{|c|}{OSF-Kvasir} & \multicolumn{1}{|c|}{FS-KID} & \multicolumn{1}{|c|}{FS-KID2} & \multicolumn{3}{|c|}{CAD-CAP} \bigstrut\\
\cline{2-7}    \multicolumn{1}{|c|}{} &   &   & \multicolumn{1}{c|}{} & \multicolumn{1}{|c|}{1\%} & \multicolumn{1}{|c|}{10\%} & 100\% \bigstrut\\

\cline{2-7}

    Random R50 & 56.8 & 24.5 & 51.6 & 33.3 & 38.5 & 71.4 \bigstrut\\
   INet Pretrained R50 & \textbf{67.0} & 44.9 & 66.1 & 39.6 & \textbf{79.3 } & \textbf{93.2} \bigstrut\\
   PIRL  & 61.8 & 40.8 & 62.2 & 29.6 & 50.7 & 88.7 \bigstrut\\
      \hline
    Ours PGCon & \textbf{66.0} & 44.9 & \textbf{68.3} & \textbf{52.9} & 75.1 & 91.01 \bigstrut\\
  Ours WINCon & 63.3 & \textbf{45.2} & 62.7 & 47.0 & 76.2 & \textbf{92.0} \bigstrut\\
    \hline
    \end{tabular}%

  \caption{Top-1 Linear n-way Classification Accuracy (n being the number of classes) evaluated with a linear classification head (2-$fc$ layers) over frozen encoder weights on datasets OSF-Kvasir, FS-KID, FS-KID2 and CAD-CAP. All results are with the R50 encoder.}
   \label{tab:LE}%
\end{table}%

\subsection{Task 3: Full Fine Tuning }
\textit{Setup:} Next, we evaluate the performance with full fine-tuning on OSF-Kvasir and CAD-CAP (sample sufficient datasets), all layers including batch-norm are trained. \\
\textit{Observations:} As seen in Table \ref{tab:fft}, PGCon closely matches or surpasses the baselines on both datasets, with higher pathology sensitivity, while WINCon only outperforms PIRL and matches the ImageNet pretraining on CAD-CAP.

\begin{table*}[htbp]

 \renewcommand*{\arraystretch}{1}
 \resizebox{1\textwidth}{!}{
  \centering
    \begin{tabular}{|c|cccc|cccc|}
    \hline
      & \multicolumn{4}{c|}{OSF-Kvasir} & \multicolumn{4}{c|}{CAD-CAP} \bigstrut\\
\cline{2-9}      & \multicolumn{1}{c|}{Top-1 Accuracy} & \multicolumn{1}{c|}{AUC} & \multicolumn{1}{c|}{Precision } & \multicolumn{1}{c|}{Recall} & \multicolumn{1}{c|}{Top-1 Accuracy} & \multicolumn{1}{c|}{AUC} & \multicolumn{1}{c|}{Precision } & \multicolumn{1}{c|}{Recall} \bigstrut\\
\hline
   Random R50  & 56.6 & 87.0 & 61.0 & 55.0 & 71.4 &  84.0 & 75.0  & 71.0 \bigstrut\\
   ImageNet Pretrained R50 & \textbf{66.6} & \textbf{88.0} & 68.0 & \textbf{67.0} & 93.1 &  98.0 & 94.0  & 94.0 \bigstrut\\
   DenseNet-161 \cite{valerio2019GIANA}  &  - &  - &  - &  - & 93.3 & 87.0 & 94.0 & 93.0 \bigstrut\\
    PIRL \cite{misra2020self} & 59.3 & 84.0 & 75.0 & 59.0 & 91.1 & 97.0 & 91.0 & 90.0 \bigstrut\\
     Guo et al. \cite{guo20203class} &  - &  - &  - &  -  & 	93.17 & - & - & -
      \bigstrut\\
    \hline
   (Ours) PGCon & \textbf{66.2} & 85.0 & \textbf{69.0} & \textbf{66.0} & \textbf{94.7} & \textbf{99.0} & \textbf{95.0} & \textbf{95.0} \bigstrut\\
    (Ours) WINCon & 62.3 & 81.0 & \textbf{69.0} & 61.0 & 93.7 & \textbf{99.0} & \textbf{95.0} & 94.0 \bigstrut\\
    \hline
    \end{tabular}%
    }
    \caption{Pathology Classification with full fine tuning on OSF-Kvasir and CAD-CAP dataset. All methods except DenseNet-161 are on ResNet-50 (R50) and full fine tuning includes BatchNorm training. The baseline ImageNet Pretrained R50 refers to supervised ImageNet pretraining with full fine-tuning. CAD-CAP : Both PGCon and WINCon match or outperform ImageNet pretraining as well as fully supervised learning  (with Imagenet pretraining) on DenseNet-161 \cite{valerio2019GIANA} , OSF-Kvasir : both methods close the gap with baselines.}
    \label{tab:fft}
\end{table*}%

\subsection{Embedding space analysis } 
\label{sec:embeddingsec} Recently, Wang et al. \cite{wang2020hyperspehere} proposed alignment and uniformity as two metrics optimized by contrastive loss in the limit of infinite negatives. We apply these metrics to evaluate the alignment (distance between samples of the same category) and uniformity (total information preservation) of our encoders. In Fig. \ref{ScoreCam}(a.) we see that PIRL exhibits high uniformity with low alignment, where high uniformity encoders are those that are highly informative of many (ideally all) the different features (less selective to some features). This is in line with the initial expectation, as PIRL does not prefer any factors over others in a dataset. PGCon, which is selective in factors, shows improved alignment but slightly reduced uniformity compared to PIRL, meaning that not all features are considered equally important to be preserved. We conjecture that the improved alignment in PGCon comes at the cost of slightly reduced uniformity and the higher performance of PGCon suggests that uniformity may be beneficial only as long as the preserved information is task relevant. The missing information may be pertaining to domain factors unnecessary for the task. We see the activations in Fig. \ref{ScoreCam}(b) supplementing this, with PIRL exhibiting spurious activations whereas PGCon shows more localized activations. 

Interestingly, WINCon is comparable to PIRL in alignment, but exhibits an even lower uniformity than PGCon, this could be due to WIN-contrast leading to further invariances in the domain. This observation is also supported by the smaller activation maps for WINCon compared to PGCon in Fig. \ref{ScoreCam}(b), as the overall uniformity decreases, the activations become more local. Furthermore, the performance indicates that WINCon is almost borderline beneficial in terms of domain invariance, we suspect it is possible to go too far and plan to investigate this aspect in future work.

\begin{figure}[!htbp]
\centering
\includegraphics[width=16cm,height=5.2cm]{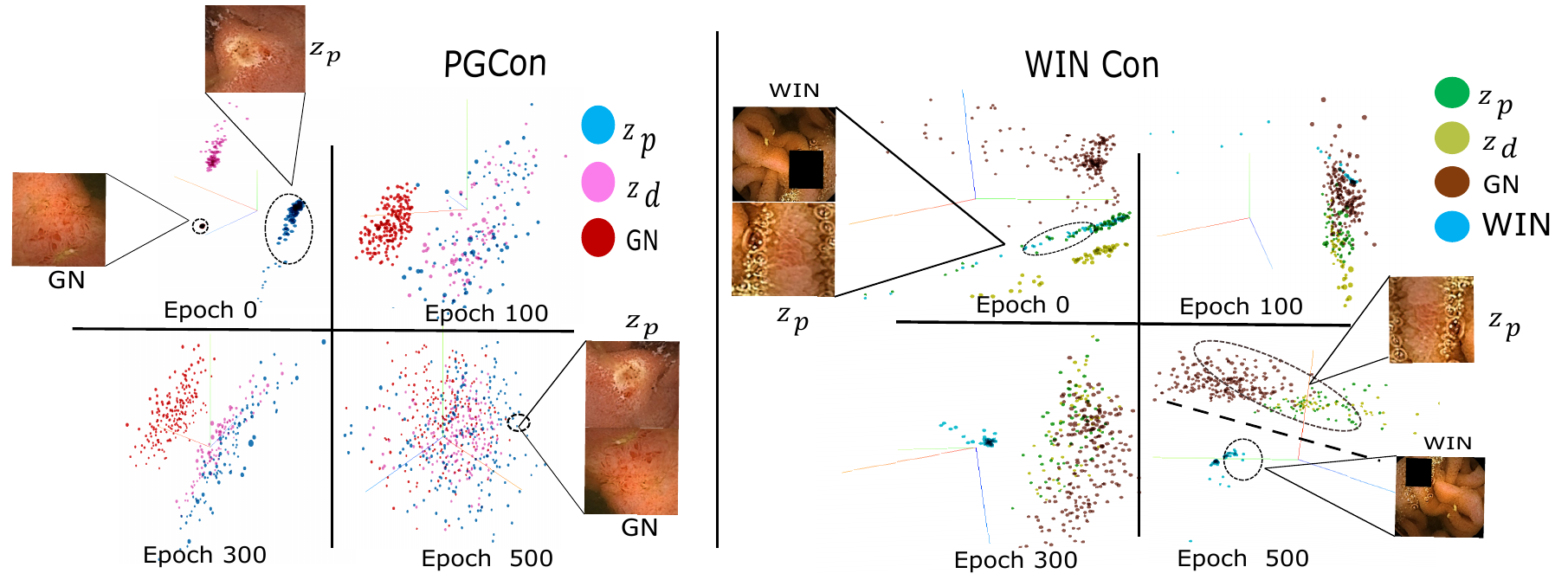}
\caption{Evolution of the embedding space: PCA of 128d feature vectors for \(z_p\), \(z_d\), Global Negatives (GN) and WIN. PGCon : Initially the embeddings start as separate localized clusters corresponding to \(z_p\), \(z_d\) and global negatives, but as the embeddings slowly specialize in pathology regions (start exhibiting invariance to other factors), the embeddings are seen to merge and spread out to a space of pathologies. WINCon : Initially the WINs lie close to corresponding prior views ($z_p$) due to being parts of the same image. However, interestingly, as the embeddings get more and more specialized in pathologies, the same WINs are pushed away from pathology based embeddings i.e., $z_p$ and GNs.  Despite the WINs being very diverse, a dense WIN cluster suggests a tendency towards  invariance to normal variations and high variance towards pathologies.}
\label{evolution}
\end{figure}

Figure \ref{evolution} shows screenshots of the feature space as it evolves during training, for PGCon and WINCon (we recommend zooming in for clarity). Recall here that the Global Negatives (GN) for an image in PGCon are prior-views from other images. The contrastive objective therefore starts by pushing apart GNs from prior views $z_p$ (Epoch 0). Over time, as embeddings specialize in representing pathologies, contrast in the feature space (between $z_p$ and GN) amounts to contrasting between specialized pathology embeddings. Epoch 500 shows the final space where instances lie based on pathological similarity (indicated by good clustering performance (Table \ref{tab:knn})).

On the other hand, we see in WINCon that initially WINs lie close to the corresponding prior views ($z_p$) as the training begins (Epoch 0), this is expected as they are from the same image. However, as training matures (epoch 500), they push away from the prior views ($z_p$). This corresponds to the desirable scenario from Fig. \ref{Views}. This is because WINs  encode normal variations, whereas ($z_p$) are pathologies, further it is seen that ($z_p$s) approach proximity with GN, which are also pathologies. 
\subsection*{Weak Labels}
\begin{figure}[!htbp]
\center
\includegraphics[]{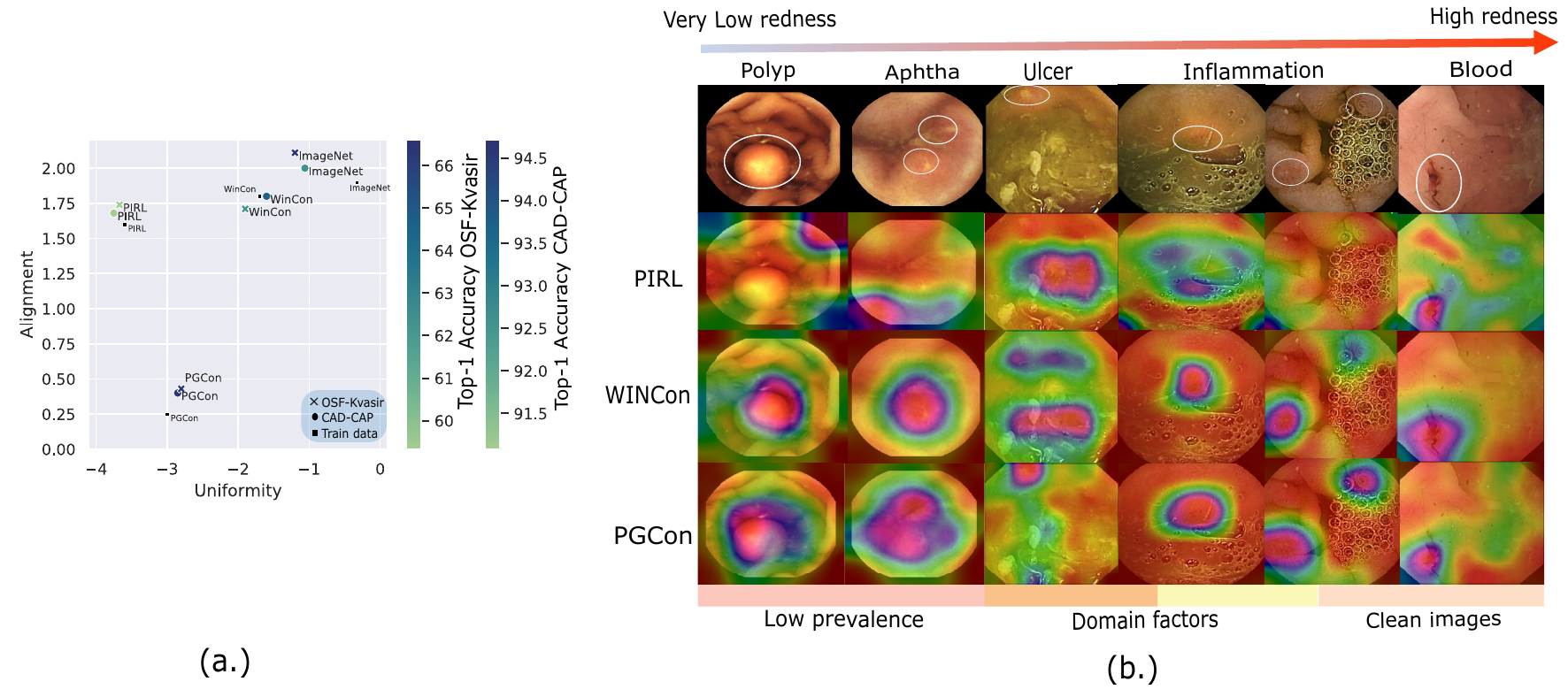}
\caption{ (a.) Alignment and Uniformity : The graph illustrates \(\mathcal{L}_{align}\) vs \(\mathcal{L}_{uniform}\) for different encoders evaluated for OSF-Kvasir, CAD-CAP and subset of train data. The points in the plot are color coded for full fine tuning accuracy. (b.) Activation map : Visualization using ScoreCAM \cite{wang2020scorecam} shows the effectiveness of our approach for non-red (polyp, ulcer) and low prevalence (in train and test sets) pathologies. It also shows that WINCon exhibits higher locality compared to PGCon. For more visualizations refer to the supplementary video.}
\label{ScoreCam}
\end{figure}

In Fig. \ref{ScoreCam} we demonstrate the generalizability of the proposed approach on a wide range of WCE pathologies.  We use ScoreCam \cite{wang2020scorecam} to visualize the activation maps on red as well as non-red pathologies, that vary in structural and other visual characteristics (for more visualizations refer to our supplementary video). Our method is also more robust to domain distractors as seen in relatively unclean images (rows 2-4). We see that WINCon activations are more local, an effect arising from contrast with normal regions within the instance, whereas PGCon activations being more robust.  Apart from the contributions discussed already, we believe this work to be of significant contribution in systematically arriving at weak labels (Fig. \ref{ScoreCam}) without any supervision on a variety of intestinal pathologies, the use of which would benefit un/semi-supervised learning in the field. 

\section{Conclusion}
In this paper, we investigate a somewhat overlooked aspect of learning under multiple-factors in datasets and discuss the challenge in preferential learning from such multi-factor datasets in a practical context, completely unsupervised. We also introduce a methodology for exploiting domain priors to guide such preferences. We hope our benchmark promotes for improved generalization, with more multi-pathology as well as multi-modality comparisons in the future, as opposed to single pathology benchmarks that limit practicality in real clinical scenarios. Our pretrained weights can be utilized for weak labeling of many different types of WCE patholgies for guiding un/semi-supervised algorithms for diagnosis.\\
\textbf{Limitations and future work :} The term generalization can be interpreted in a variety of ways from one domain to another. In WCE classification, first-stage generalization applies to counteracting the domain shift arising from using different capsule endoscopes, different organs (small bowel, colon, esophagus) as well as to different pathologies in these environments.  In this paper we test generalization with respect to capsule modality as well as pathologies (of small bowel and colon), however, in the future, we plan to expand generalization outside this premise and expand the generalization to tasks like localization, organ classification etc. Furthermore, we aim to investigate robust-priors that can be attuned to different medical domains, to broaden the scope of such preferential learning. 
\section*{Data availability}
The datasets OSF Kvasir, KID, KID2 are publicly available, datasets CAD-CAP, Ps-DeVCEM may be available from the original sources upon reasonable request.

\bibliography{sample}


\section*{Author contributions statement}

A.V. conceived the study, A.V., M.A. designed the experiments, A.V conducted the experiments. All authors analysed the results. A.V. wrote the manuscript and all authors reviewed and revised the manuscript. 

\section*{Comping interests}
The authors declare no competing interests.

\end{document}